\documentclass[conference]{IEEEtran}

\usepackage{cite}
\usepackage{amssymb}
\usepackage{amsmath}
\usepackage{graphicx}
\usepackage{enumitem}
\usepackage{caption}
\usepackage{subcaption}
\usepackage{multirow}

\ifCLASSINFOpdf

\else
 
\fi

\begin{document}

\title{Transient Analysis of a Resource-limited Recovery Policy for Epidemics: a Retrial Queueing Approach}



%
\author{\IEEEauthorblockN{Aresh Dadlani\IEEEauthorrefmark{1},
Muthukrishnan Senthil Kumar\IEEEauthorrefmark{2},
Kiseon Kim\IEEEauthorrefmark{1} and
Faryad Darabi Sahneh\IEEEauthorrefmark{3}}
\IEEEauthorblockA{\IEEEauthorrefmark{1}School of EECS,
Gwangju Institute of Science and Technology, Gwangju 61005, South Korea}
\IEEEauthorblockA{\IEEEauthorrefmark{2}Department of Applied Mathematics and Computational Science, PSG College of Technology, India}
\IEEEauthorblockA{\IEEEauthorrefmark{3}School of Computer Science, Georgia Institute of Technology, Atlanta, GA 30332, USA\\
Email: dadlani@gist.ac.kr, msk@amc.psgtech.ac.in, kskim@gist.ac.kr, fsahneh3@gatech.edu}}

\maketitle

\begin{abstract}
Knowledge on the dynamics of standard epidemic models and their variants over complex networks has been well-established primarily in the stationary regime, with relatively little light shed on their transient behavior. In this paper, we analyze the transient characteristics of the classical susceptible-infected (SI) process with a recovery policy modeled as a state-dependent retrial queueing system in which arriving infected nodes, upon finding all the limited number of recovery units busy, join a virtual buffer and try persistently for service in order to regain susceptibility. In particular, we formulate the stochastic SI epidemic model with added retrial phenomenon as a finite continuous-time Markov chain (CTMC) and derive the Laplace transforms of the underlying transient state probability distributions and corresponding moments for a closed population of size $N$ driven by homogeneous and heterogeneous contacts. Our numerical results reveal the strong influence of infection heterogeneity and retrial frequency on the transient behavior of the model for various performance measures.
\end{abstract}


%
\IEEEpeerreviewmaketitle

\section{Introduction}
Epidemiological models have assumed new relevance in assessing spreading processes over a broad interdisciplinary spectrum.  Ranging from modeling malware propagation on the digital landscape \cite{Mieghem2009, Dadlani2014} to `word-of-mouth' influence in social networking platforms \cite{Hill2010, Wei2013}, analysis of classical stochastic epidemic models and their deterministic approximations at microscopic and macroscopic levels have been the subject of serious scientific inquiry in recent years \cite{Nowzari2016}. Much of the existing works however, are mostly concerned with the long-term characteristics of the epidemics rather than their transient behavior upto some specified time. In fact, time-dependent analysis provides deeper insight on the system behavior when the primary parameters are perturbed and thus, can serve crucial in devising effective control measures.

In regard to applications of queueing theory in quantitative analysis of epidemic progression, the authors of \cite{Trapman2009} showed the number of infected nodes at the moment of first detection to be geometrically distributed by formulating the susceptible-infected-removed (SIR) epidemic as an $M/G/1$ queue with processor sharing service discipline. Approximations for the quasi-stationary distribution (QSD) of the number of susceptibles in the susceptible-infected-susceptible (SIS) and susceptible-latent-infected-susceptible (SEIS) models were derived in \cite{Carlos2010}, wherein each node was either a busy (infected) or an idle (susceptible) server in a homogeneously mixing population. Similarly, analytical derivations of the QSD and transient distribution for the maximum number of infectives in a generalized SIS model, described as a birth-death process, were detailed in \cite{Artalejo2010}. Furthermore, to investigate the number of infectives resulting from a computer virus (CodeRed-II) attack during time interval $(0,t]$, the block-structured state-dependent event (BSDE) approach was advocated considering non-exponential and correlated infection and recovery flows in an SIS-type model  \cite{Amador2013}. In view of heterogeneous infectiousness and susceptibility in the SIS model, Economou {\it et al.} \cite{Economou2015} studied the behavior of the corresponding $2^N$-state Markov chain formulation in the quasi-stationary regime. Moreover, Sahneh {\it et al.} \cite{Sahneh2013} deduced the state occupancy probabilities of the SIS model with multiple contact layers by reducing the exact $M^N$-state Markov chain representation to a system of $MN$ non-linear differential equations using mean-field approximation. The above works nonetheless, do not consider a recovery policy, which inherently has limited resources, within their models.

In this paper, we address the impact of infected nodes retrying for recovery controlled by limited resources on the transient behavior of the SI model considering both, homogeneous and heterogeneous\footnote{Heterogeneous contact in here refers to the non-uniform infection rate associated with each node in any arbitrary network topology.} contacts in a finite population. We build our time-homogeneous CTMC model based upon the retrial queueing notion in which infected nodes return back to become susceptible only after being served by one of the idle recovery units. On finding all units busy, the infected node then joins a virtual buffer, known as orbit, from which it retries persistently for access  until granted service by an idle unit. Our main interest is to numerically investigate the transient state probability distributions of the number of infected nodes undergoing recovery and those residing in the orbit as well as their related moments under varying parametric settings which, to the best of our knowledge, has not yet been reported.

An immediate practical application of our work is the time and workload dependency analysis of online computer scanning services. Under a practically relevant scenario comprising of a small number of networked devices, an infected client device attempts to access a server that provides malware detection/eradication services. In reality, the number of communication ports on the server dedicated for this purpose is always limited and thus, cannot serve all clients simultaneously. On arrival, if a client finds an idle port on the server, it connects successfully to receive the required service. Otherwise, the client retries randomly and independently, expecting to gain access to an idle port on the server. Therefore, for a finite operation time horizon, statistical information on the proportion of infected clients being scanned and those awaiting access serves substantial to network administrators.

The remainder of this paper is organized as follows. The Markovian retrial SI model with limited recovery resources is formally introduced in Section \ref{sec2}. Derivations for the transient state and marginal probabilities under different contact types are provided in Section \ref{sec3}, followed by numerical results in Section \ref{sec4}. Finally, we  conclude the paper in Section \ref{sec5} with directions for potential future work.

\section{Proposed Model Description}
\label{sec2}
To make the subsequent derivations systematic, we introduce some graph-theoretical nomenclature and the Markov chain representation of the retrial SI model in this section.

\subsection{Contact Network Topology}
Consider a fixed network of size $N$ within which a particular infection spreads. We represent such a contact network as an undirected graph $G=(V,E)$, where $V=\{1,2,\ldots,N\}$ denotes the set of constituent nodes and $E \subseteq V \times V$ is the set of interaction links. The associated adjacency matrix of $G$ is given as $A \triangleq [a_{i,j}]_{V \times V}$, where $a_{i,j} = 1$ if $i$ and $j$ are adjacent neighbors in contact, and $a_{i,j} = 0$ if otherwise. Following the definition of matrix $A$, the degree of any node $i \in V$ can be easily computed to be $d_i=\sum_{j=1}^N a_{i,j}$.

\subsection{Retrial SI Model Formulation}
\begin{figure}[!t]
\centering
\includegraphics[width=3.2in]{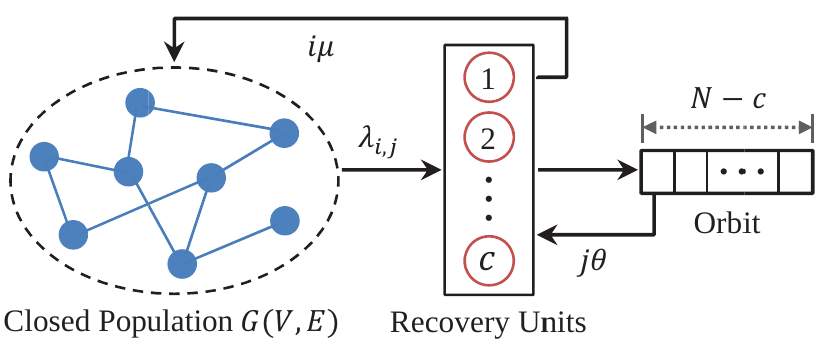}
\caption{Schema of the state-dependent retrial SI model for network of size $|V|\!=\!N$ with finite recovery units ($c\!<\!N$).}
\vspace{-0.5em}
\label{fig1}
\end{figure}
We extend the standard stochastic SI compartmental model by reinforcing the intrinsic retrial behavior of infected nodes contending for limited treatment resources as shown in Fig.~\ref{fig1}. Specifically, each node transitions from being in the susceptible (\textbf{S}) sub-population to the infected (\textbf{I}) sub-population, and back again to susceptible upon receiving treatment. Since the population is closed, i.e. $|\textbf{S}|\!+\!|\textbf{I}|\!=\!N$, the system state at time $t\!\geq\!0$ can be fully described by $I(t)$ and $R(t)$ which represent the number of recovery units $(c)$ being occupied by infected nodes and the number of retrying infected nodes in the orbit of size $N\!-\!c$, respectively. The arrival of infected nodes is assumed to follow a Poisson process with state-dependent arrival rate of $\lambda_{i,j} \!\in\! \mathbb{R}^+$, where $i\!\in\!\{0,1,\ldots,c\}$ and $j\!\in\!\{0,1,\ldots,N\!-\!c\}$. Also, an arriving infected node undergoes recovery at one of the units for an exponentially distributed time with mean $\mu^{-1}\! \in\! \mathbb{R}^+$, after which it once again becomes susceptible to the contagion. Infected nodes awaiting in the orbit attempt for service at exponentially distributed random time intervals with mean $\theta^{-1}\!\in \mathbb{R}^+$. Following these definitions, the finite CTMC representation of the retrial SI epidemic can be described as the following bi-variate process:
\begin{equation}
\label{eq1}
	X(t) = \big\{\big(I(t),R(t)\big); t \geq 0\big\},
\end{equation}
taking values on state space $\Omega\!=\!\big\{(i,j)|i\!\in\!\{0,1,\ldots,c\}; j\!\in\!\{0,1,\ldots,N\!-\!c\}\big\}$. For any arbitrary $x,y \in \Omega$ with $x=(i,j)$ indicating the state of the system having $i$ recovery units busy treating infected nodes and $j$ number of infected nodes in the orbit at time $t$, the infinitesimal transition probabilities are specified as follows, where $q_{x,y}\in \mathbb{R}_{\geq0}$ denotes the transition rate from state $x$ to state $y$:
\begin{equation} 
	\label{eq2}
	\begin{split}
		P_{x,y}(t,t+\Delta t) & \triangleq\text{Pr}\big[X(t+\Delta t)=y | X(t)=x\big] \\
 							  & =\! \begin{cases}
   								 q_{x,y}\Delta t+o(\Delta t), & \text{if } x \neq y \\
    							 0, & \text{if otherwise. } 
  									\end{cases}
	\end{split}
\end{equation}

For complete characterization of the time evolution of $X(t)$, we define $\mathbb{P}(t)\!=\![p_{i,j}(t)]_{1 \times |\Omega|}$ to be the transient state probability vector, with element $p_{i,j}(t)$ denoting the probability of process $X(t)$ being in state $(i,j)$ at time $t$, i.e. $\forall (i,j) \in \Omega$:
\begin{equation}
\label{eq3}
	p_{i,j}(t) \triangleq \text{Pr}[X(t)\!=\!(i,j)] = \text{Pr}[I(t)\!=\!i \text{ and } R(t) = j].
\end{equation}

\section{Transient State Analysis}
\label{sec3}
The nature of node-level interactions has been shown to profoundly impact the process of contagion and its control mechanisms \cite{Sahneh2011, Preciado2013, Sahneh2013}. In this section, we derive the transient solution of the state occupancy probabilities.

\subsection{Retrial SI Model with Homogeneous Contacts}
\begin{figure}[!t]
\centering
\includegraphics[width=3.45in]{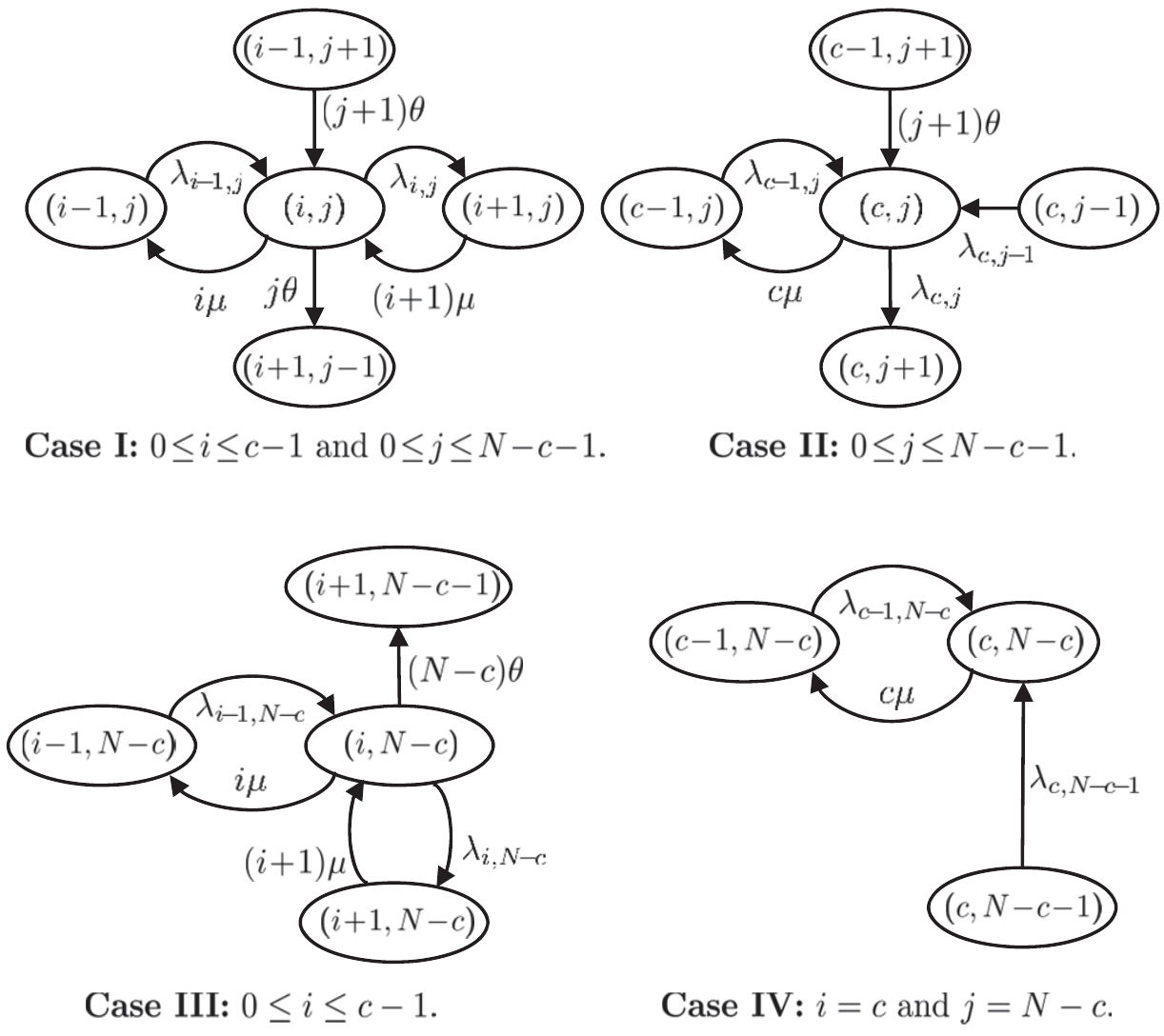}
\caption{State transitions of the stochastic retrial SI model.}
\vspace{-0.5em}
\label{fig2}
\end{figure}

In this setting, we assume the spread of a typical infection to be driven by homogeneous mixing in a population wherein each node makes contact with another node at random time intervals which are i.i.d. random variables \cite{Carlos2010}. Subsequently, the state-dependent arrival rate of infected nodes at the recovery units is expressed as $\lambda_{i,j}=\alpha (N-i-j)/N$, where $\alpha$ is the contact rate in the population. With the time-homogeneous process $X(t)$ defined over $\Omega$ of size $(c\!+\!1)(N\!-\!c\!+\!1)$, the transition rates dictating the retrial SI model dynamics are:
\begin{equation}
\label{eq4}
	q_{x,y}\! =\! \left.
  \begin{cases}
    \lambda_{i,j}, & \text{if } y\!=\!(i\!+\!1,j);\; i\!\leq\!c\!-\! 1, j\!\leq\!N\!-\!c \\
    i\mu, & \text{if } y\!=\!(i\!-\!1,j);\; 1\!\leq\! i \!\leq\! c, j \!\leq\! N\!-\!c \\
    j\theta, & \text{if } y\!=\!(i\!+\!1,j\!-\!1);\; i\!\leq\!c\!-\!1, 1\!\leq\!j\!\leq\!N\!-\!c \\
    \lambda_{c,j}, & \text{if } y\!=\!(c,j\!+\!1);\; j\!\leq\!N\!-\!c\!-\!1 \\
    0, & \text{if otherwise. } 
  \end{cases}
  \right.
\end{equation}
Equation (\ref{eq4}) simply expresses the four possible cases of state transitions shown in Fig.~\ref{fig2}. The corresponding Chapman-Kolmogorov forward differential equations are:
\begin{itemize}[leftmargin=*]
	\item \textbf{Case I:} When the set of recovery units ($0 \leq i \leq c-1$) and the orbit ($0 \leq j \leq N-c-1$) have vacancies:
\begin{equation} 
	\label{eq5}
	\begin{split}
		p'_{i,j}(t) = & -\!(\lambda_{i,j}\!+\!i\mu\!+\!j\theta)p_{i,j}(t)\!+\!\lambda_{i\!-\!1,j}p_{i\!-\!1,j}(t) \\
 							  & +\!(j\!+\!1)\theta p_{i\!-\!1,j\!+\!1}(t)\!+\!(i\!+\!1)\mu p_{i\!+\!1,j}(t).
	\end{split}
\end{equation}

	\item \textbf{Case II:} When all the recovery units are occupied ($i=c$) and the orbit is not full ($0 \leq j \leq N-c-1$):
\begin{equation} 
	\label{eq6}
	\begin{split}
		p'_{c,j}(t) = & -\!(\lambda_{c,j}\!+\!c\mu)p_{c,j}(t)\!+\!\lambda_{c\!-\!1,j}p_{c\!-\!1,j}(t) \\
 							  & +\!(j\!+\!1)\theta p_{c\!-\!1,j\!+\!1}(t)\!+\!\lambda_{c,j\!-\!1} p_{c,j\!-\!1}(t).
	\end{split}
\end{equation}

	\item \textbf{Case III:} When at least one recovery unit is idle ($0 \leq i \leq c-1$) and the orbit is full ($j=N-c$):
\begin{equation} 
	\label{eq7}
	\begin{split}
		p'_{i,N\!-\!c}(t) = & -\!(\lambda_{i,N\!-\!c}\!+\!i\mu\!+\!(N\!-\!c)\theta)p_{i,N\!-\!c}(t)  \\
 							  & +\!\lambda_{i\!-\!1,N\!-\!c}p_{i\!-\!1,N\!-\!c}(t)\!+\!(i\!+\!1)\mu p_{i\!+\!1,N\!-\!c}(t).
	\end{split}
\end{equation}	
	
	\item \textbf{Case IV:} When the finite set of recovery units ($i=c$) and the orbit ($j=N-c$) are all full:
\begin{equation} 
	\label{eq8}
	\begin{split}
		p'_{c,N\!-\!c}(t) = & -\!c\mu p_{c,N\!-\!c}(t)\!+\!\lambda_{c\!-\!1,N\!-\!c} p_{c\!-\!1,N\!-\!c}(t)  \\
 							  & +\!\lambda_{c,N\!-\!c\!-\!1} p_{c,N\!-\!c\!-\!1}(t).
	\end{split}
\end{equation}	
\end{itemize}

Given $\mathbb{P}(0)$, equations (\ref{eq5})-(\ref{eq8}) can be written in the matrix form as $\mathbb{P}'(t)=\mathbb{P}(t)\cdot Q$, where $Q:\Omega\times\!\Omega \rightarrow \mathbb{R}_{\geq 0}$ is the infinitesimal generator matrix with elements $q_{x,y}$ given as:
\begin{equation}
\label{eq9}
	Q(x,y) = \left.
  \begin{cases}
    q_{x,y}, & \text{if } x \neq y \\
    -\sum\limits_{\substack{y \in \Omega \\ y\neq x}} q_{x,y}, & \text{if } x = y.    
  \end{cases}
  \right.
\end{equation}
We now employ the Laplace Transform (LT) operator $\mathcal{L}[\cdot]$ on (\ref{eq5})-(\ref{eq8}) to obtain $\mathbb{P}^*(s)=\mathcal{L}[\mathbb{P}(t)]$, which is then used to yield the probability vector $\mathbb{P}(t)$ through inverse LT. As a result, we arrive at the following system of equations:
\begin{equation} 
	\label{eq10}
	\begin{split}
		p_{i,j}(0) = & (s\!+\!\lambda_{i,j}\!+\!i\mu\!+\!j\theta)p^\ast_{i,j}(s)\!-\!\lambda_{i\!-\!1,j}p^\ast_{i\!-\!1,j}(s)  \\
 					 & -\!(j\!+\!1)\theta p^\ast_{i\!-\!1,j\!+\!1}(s)\!-\!(i\!+\!1)\mu p^\ast_{i+1,j}(s),
	\end{split}
\end{equation}	
\begin{equation} 
	\label{eq11}
	\begin{split}
		p_{c,j}(0) = & (s\!+\!\lambda_{c,j}\!+\!c\mu)p^\ast_{c,j}(s)\!-\!\lambda_{c\!-\!1,j}p^\ast_{c\!-\!1,j}(s)  \\
 					 & -\!(j\!+\!1)\theta p^\ast_{c\!-\!1,j\!+\!1}(s)\!-\!\lambda_{c,j-1} p^\ast_{c,j\!-\!1}(s),
	\end{split}
\end{equation}
\begin{equation} 
	\label{eq12}
	\begin{split}
		p_{i,N\!-\!c}(0) = & (s\!+\!\lambda_{i,N\!-\!c}\!+\!i\mu\!+\!(N\!-\!c)\theta)p^\ast_{i,N\!-\!c}(s)  \\
 					 & -\!\lambda_{i\!-\!1,N\!-\!c}p^\ast_{i\!-\!1,N\!-\!c}(s)\!-\!(i\!+\!1)\mu p^\ast_{i\!+\!1,N\!-\!c}(s),
	\end{split}
\end{equation}
\begin{equation} 
	\label{eq13}
	\begin{split}
		p_{c,N\!-\!c}(0) = & (s\!+\!c\mu)p^\ast_{c,N\!-\!c}(s)\!-\!\lambda_{c\!-\!1,N\!-\!c}p^\ast_{c\!-\!1,N\!-\!c}(s)  \\
 					 & -\!\lambda_{n,N\!-\!c\!-\!1} p^\ast_{c,N\!-\!c\!-\!1}(s).
	\end{split}
\end{equation}
Re-arranging (\ref{eq10})-(\ref{eq13}) in the vector-matrix form results in $\mathbb{P}^\ast(s)\!=\!\mathbb{P}(0)M^{-1}$, where the invertible matrix $M$ exhibits the following block tridiagonal structure:
\begin{equation}
\label{eq14}
M=
\begin{bmatrix}
    A_{0}   & B_{0}   & 0      & \dots  & 0      & 0     \\
    C_{1}   & A_{1}   & B_{1}  & \dots  & 0      & 0     \\
    0       & C_{2}   & A_{2}  & \dots  & 0      & 0     \\
    \vdots  & \vdots  & \vdots & \ddots & \vdots & \vdots\\
    0       & 0       & 0      & \dots  & C_{c}  & A_{c}
\end{bmatrix}_{(N-c+1)(c+1)}
\end{equation}
For $0\leq\!i\!\leq\!c-1$, the diagonal sub-matrix $A_i\!=\!(s+\lambda_{i,j}+i\mu+j\theta)I_{N-c+1}$, with $I$ denoting the identity matrix, whereas $A_c\triangleq[\hat{a}_{u,v}]$ is an upper bidiagonal matrix featured as:
\begin{equation}
\label{eq15}
	\hat{a}_{u,v} = \left.
  \begin{cases}
    s+\lambda_{c,j}+c\mu, & \text{if } u = v \neq N-c \\
    -\lambda_{c,j}, & \text{if } u = v - 1 \\
    s+c\mu, & \text{if }  u = v = N - c \\
    0, & \text{if otherwise. } 
  \end{cases}
  \right.
\end{equation}
where $u$ and $v$ are unique integer values returned by the labeling function $f:\Omega \rightarrow  \mathbb{N}_{\geq0}$, such that $f(i,j)=(N-c+1)i+j$. The diagonal sub-matrix $C_i\!=\!-i\mu I_{N-c+1}$ and $B_i\!\triangleq\![\hat{b}_{u,v}]$ is a lower bidiagonal matrix with entries:
\begin{equation}
\label{eq16}
	\hat{b}_{u,v} = \left.
  \begin{cases}
    s+\lambda_{i,j}, & \text{if } u = v \\
    -j\theta, 		 & \text{if } u = v + 1 \\
    0, 				 & \text{if otherwise. } 
  \end{cases}
  \right.
\end{equation}
After computing $\mathbb{P}^\ast(s)$ as above, we now find the marginal distributions of $I(t)$ and $R(t)$, denoted respectively as $p_i(t)$ and $q_j(t)$, and their moments using inverse LT. Once identified, the stationary probability vector, $\Pi=[\pi_{i,j}]_{1\times|\Omega|}$, can also be determined using $\Pi\!=\!\lim_{s\to 0}\mathbb{P}^\ast(s)\!=\!\lim_{t\to\infty}\mathbb{P}(t)$. The probability of finding $i$ infected nodes under recovery at time instant $t$ is given as below:
\begin{equation}
\label{eq17}
	p_i(t) = \sum_{j=0}^{N-c} p_{i,j}(t),
\end{equation}
with corresponding moments $E[I^n(t)]\!=\!\sum_{i=0}^c i^n p_i(t)$. Likewise, the marginal distribution of $R(t)$ is expressed as:
\begin{equation}
\label{eq18}
	q_j(t) = \sum_{i=0}^{c} p_{i,j}(t),
\end{equation}
with $E[R^n(t)]=\sum_{j=0}^{N-c} j^n q_j(t)$ as its $n^{th}$ raw moment.

\subsection{Retrial SI Model with Heterogeneous Contacts}
Unlike the homogeneous counterpart, where an infected node can equally infect any susceptible node, heterogeneity in the retrial SI model can be reflected in terms of non-homogeneous (node-dependent) infection rates over the contact network as in the works of \cite{Economou2015} and \cite{Mieghem2012}. To this end, we assume that any susceptible node, say $k$, can be infected in two ways: (i) infection stemming from some external source (outside the population) according to a Poisson process with rate $\delta^k_{i,j} \in \mathbb{R}^+$ such that $\delta^k_{i,j}=\alpha_k(N-i-j)/N$ and (ii) internal nodal infection following a Poisson process with rate $\beta_{k,l} \in \mathbb{R}^+$ on $k$ for all susceptible nodes $l$ such that $a_{k,l}\!=\!1$. Without loss of generality, all involved processes governing the external/internal infections, recovery, and retrial times are assumed to be mutually independent. This scenario is analogous to virus propagation in computer networks, where node $k$ not only receives the virus from its neighbors, but can also generate and spread its own virus. For convenience, we denote the state-dependent arrival rate due to node $k$ as follows, where \textbf{I} symbolizes the sub-population of infected nodes in the network:
\begin{equation}
\label{19}
	\lambda^k_{i,j} = \delta^k_{i,j}+\sum_{l\in \text{\textbf{I}}}\beta_{l,k} a_{l,k}.
\end{equation}

By replacing $\lambda_{i,j}$ and $\lambda_{c,j}$ in (\ref{eq4}) with $\lambda^k_{i,j}$ and $\lambda^k_{c,j}$, respectively, the state transition rates for the proposed model with heterogeneities can be expressed in a similar manner. Repetition of the expressions for the Kolmogorov forward equations and operations in the LT domain are not included here due to limited space. Thus, undertaking the same approach as in the preceding sub-section, with $p^k_{i,j}(t)$ defined as the probability of having $i$ recovering and $j$ orbital infected nodes due to $k$ at time $t$ (and the other notations varied accordingly), the respective marginal distributions and moments can be obtained in a straightforward manner.

\section{Numerical Results and Discussions}
\label{sec4}
The objective of resorting to numerical simulations in this section is to analyze the transient state behavior of the retrial SI model with respect to the derived performance measures under varying parametric values so as to visualize the solutions in practical scenarios. Since inverting the computed LT is quite tedious, particularly for larger values of $N$ and $c$, we adopt the Jagerman-Stehfest method which is built upon the Post-Widder inversion formula to numerically compute the approximate results for $\mathcal{L}^{-\!1}[\mathbb{P}^\ast(s)]$ and $\mathcal{L}^{-\!1}[\mathbb{P}^{k\ast}(s)]$ \cite{Abate1995}. Throughout this section, we assume the initial condition of the system to be $p_{0,0}(0)\!=\!1$ and $k=2$.

\subsection{Marginal Distributions at Different Time Points}
\begin{figure}[!t]
    \centering
    \begin{subfigure}{0.46\textwidth}
        \includegraphics[width=\textwidth]{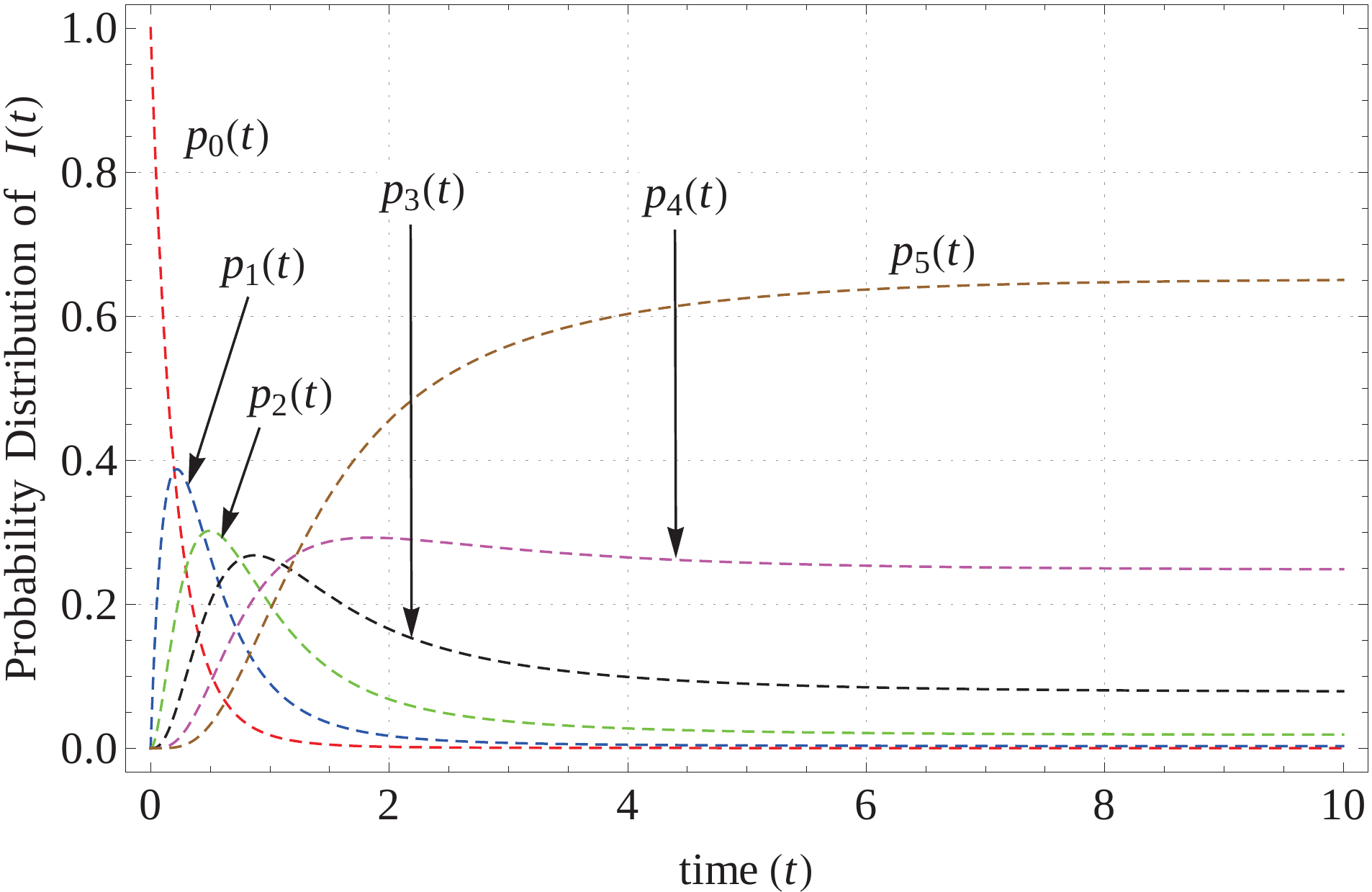}
        \caption{Marginal distribution of $I(t)$.}
        \label{fig3a}
        \vspace{0.5em}
    \end{subfigure}
      
    \begin{subfigure}{0.46\textwidth}
        \includegraphics[width=\textwidth]{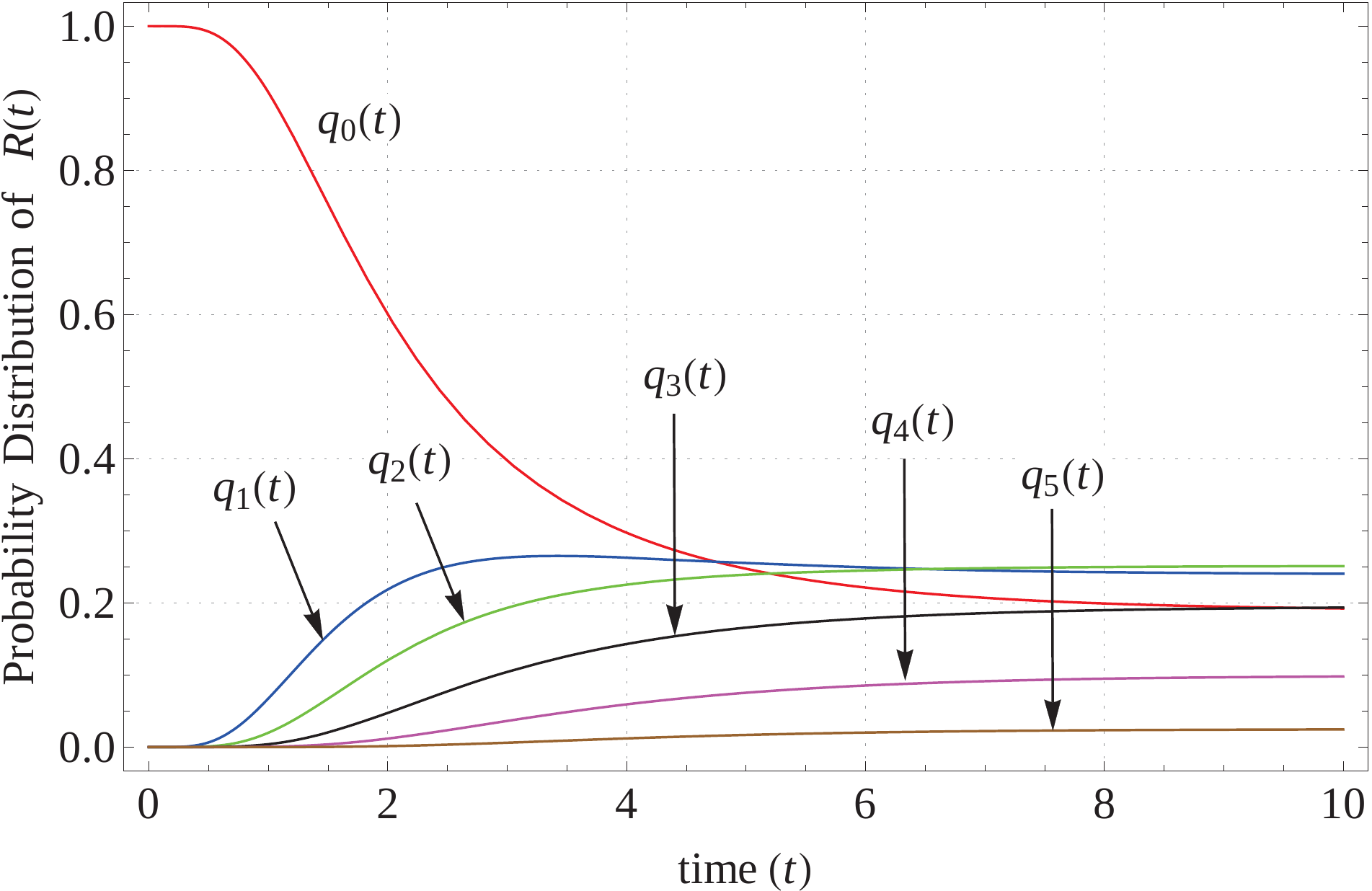}
        \caption{Marginal distribution of $R(t)$.}
        \label{fig3b}
    \end{subfigure}
    \caption{Transient behavior under homogeneous contacts for $\!N\!=\!10$, $\!c\!=\!5$, $\alpha\!=\!5$, $\!\mu\!=\!0.4$, $\theta\!=\!2$, and $p_{0,0}(0)\!=\!1$.}
    \vspace{-0.5em}
    \label{fig3}
\end{figure}

In Fig.~\ref{fig3}, the marginal probabilities $p_i(t)$ and $q_j(t)$ for a well-mixed population are plotted with parameters $(N,c)\!=\!(10,5)$, $\alpha\!=\!5$, $\mu\!=\!0.4$, and $\theta\!=\!2$. Under this parametric set-up, we observe in Fig.~\ref{fig3a} that within time interval $t\!\in\![0,2]$, unlike the rapid exponential decay of $p_0(t)$, the probability of having at least one infected node under recovery gradually rises before approaching the steady-state probability. This clearly indicates that with the infected nodes initially arriving with maximum rate $\lambda_{0,0}\!=\!\alpha$, the probability of the recovery units being occupied increases and eventually, the probability of having all units busy ($p_5(t)$) reaches its highest in long-term with the least probability for them being all idle. On the other hand, Fig.~\ref{fig3b} shows that $q_0(t)=1$ upto $t=0.5$ and then reduces to its minimum value at a slower rate. Dependent on the initial conditions, such behavior in interval $t\in[0,9]$ is not far from expectation. As the infected nodes begin to arrive, they are immediately served by the idle units and enter the orbit only if all units are busy. Thus, decrease in $q_0(t)$ reflects the recovery units' availability for service and as the number of arrivals outnumber the recovery units, the orbit gradually begins to fill up thus, reducing the  probability of it being vacant.

\begin{figure}[!t]
    \centering
    \begin{subfigure}{0.46\textwidth}
        \includegraphics[width=\textwidth]{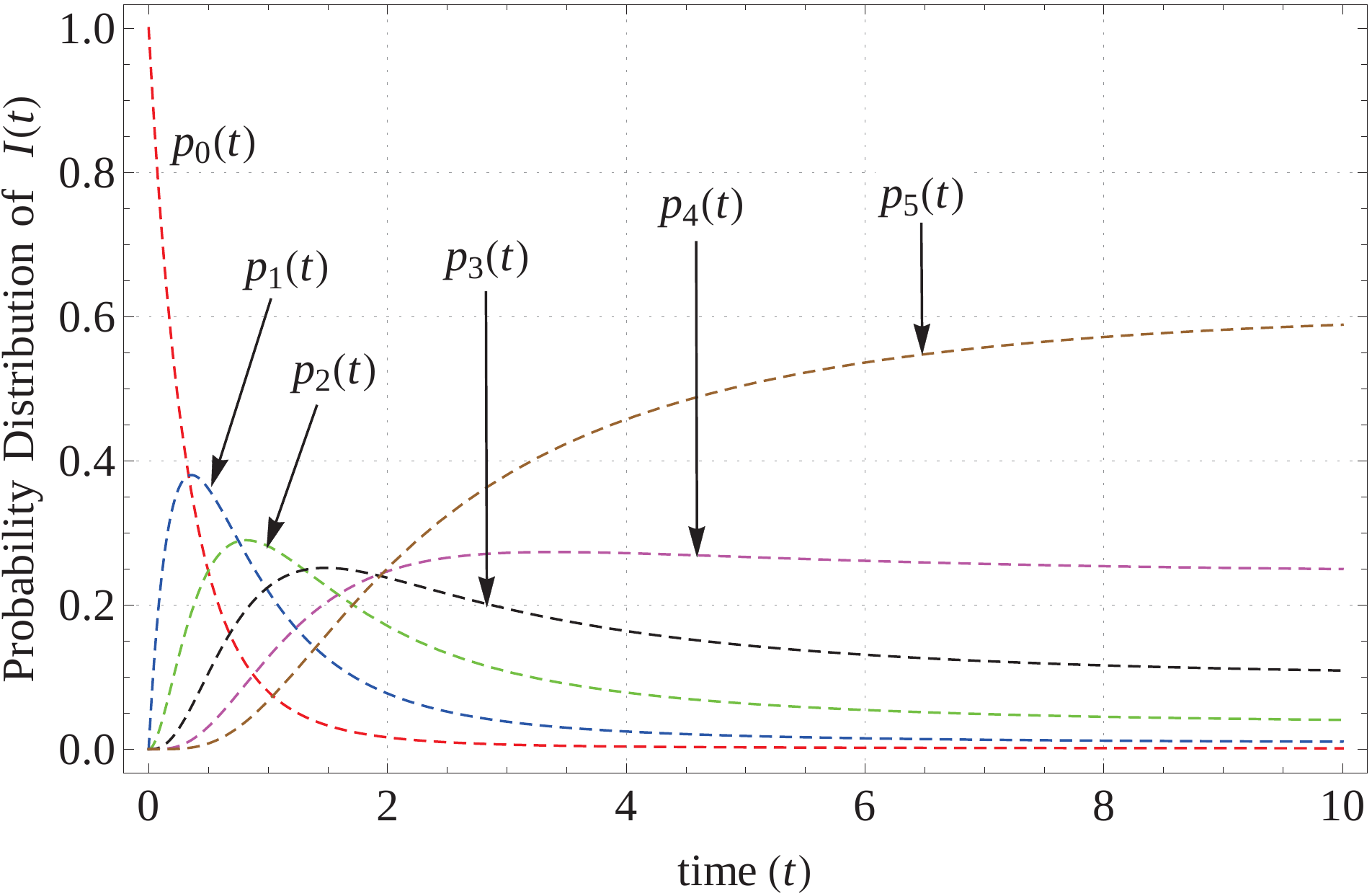}
        \caption{Marginal distribution of $I(t)$.}
        \label{fig4a}
        \vspace{0.5em}
    \end{subfigure}
      
    \begin{subfigure}{0.46\textwidth}
        \includegraphics[width=\textwidth]{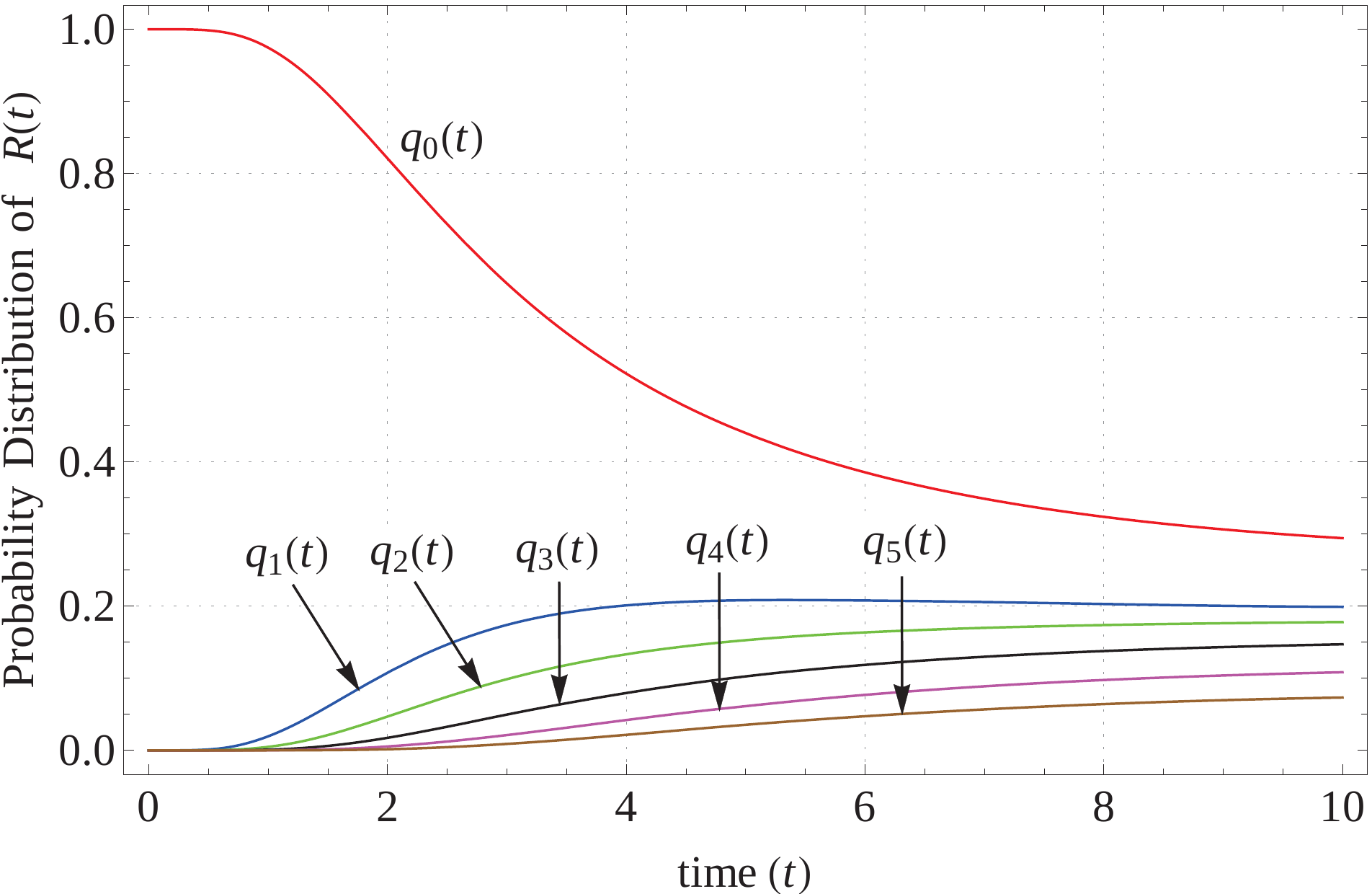}
        \caption{Marginal distribution of $R(t)$.}
        \label{fig4b}
    \end{subfigure}
    \caption{Transient behavior under heterogeneous contacts for $N\!=\!10$, $c\!=\!5$, $\alpha\!=\!5$, $\mu\!=\!0.4$, $\theta\!=\!2$, $\beta_{k,l}\!=\!1$, and $p_{0,0}(0)\!=\!1$.}
    \vspace{-0.5em}
    \label{fig4}
\end{figure}

Marginal distributions for the same system under infection heterogeneity are illustrated in Fig.~\ref{fig4}, where $\alpha_k\!=\!\alpha d_k/N$ and the internal infection spreading rate from node $k$ to its susceptible neighbors is $\beta_{k,l}\!=\!d_l/N$. The relatively longer transient phase ($t\in[0,14]$) in this figure is evidence of the impact of non-uniform infection over the contact network. In other words, since $\alpha_k$ and $\beta_{k,l}$ are node degree-bounded rates, the infection spread is less spontaneous thus, resulting in a longer time for healthy nodes to get infected in comparison to uniform infectivity in Fig.~\ref{fig3}. Hence, a priori information on the connectivity pattern of the network is required in order to control the spread under heterogeneous infectivities.

\subsection{Expected Number of Infectives at Different Time Points}
Tables~\ref{table1} and \ref{table2} summarize the expected number of infected nodes under recovery and awaiting treatment in the orbit at discrete time epochs for the two contact types. A common trend visible in both models is that the expected number of infectives under recovery increases with the value of $c$ for any arbitrary population size. Increasing $c$ improves the chances of finding an idle recovery unit which in turn, reduces the average number of infected nodes in the orbit. The transient behavior exhibited by the retrial SI model with heterogeneous infection rates however, exists for a comparatively longer time interval before reaching stationarity as shown in Table~\ref{table2}. For instance, the average number of orbital nodes for $(N,c)\!=\!(20,15)$ stabilizes at $t\!=\!10$ in Table~\ref{table1}, while heterogeneity in the infection rate prolongs this convergence to $t\!>\!20$ as shown in Table~\ref{table2}. It should also be noted that the system never approaches an infection-free equilibrium (where all nodes are susceptible) due to the state dependency of the infection rate. Therefore, $\lambda_{i,j}$ reaches its maximum value when $(i,j)\!=\!(0,0)$ and becomes zero when all the nodes are infected, i.e. $(i,j)\!=\!(c,N-c)$.
\begin{table}[!t]
\caption{First moments under homogeneous contacts.}
\vspace{-0.5em}
\begin{center}\small
\begin{tabular}{|l|l|c|c|c|}
\hline
\multicolumn{2}{|c|}{\!\big($E[I(t)]$, $E[R(t)]$\big)\!}  	        &  $N\!=\!10$   &	$N\!=\!20$    &   $N\!=\!40$      \\ \hline\hline
\multirow{4}{*}{\!$c\!=\!5$\!}	& $t\!=\!0.5$		&  (2.81,1)		&	(2.79,1)      &	  (2.78,1.03)  \\ 
 						        & $t\!=\!2$		    &  (2.36,1.65)  &	(1.79,2.52)   &	  (1.53,3.14)  \\
 						        & $t\!=\!5$	    	&  (1.72,2.52)  &	(0.93,5.64)   &	  (0.63,8.46)  \\
 								& $t\!=\!10$		&  (1.62,2.69)  &   (0.69,7.62)   &	  (0.37,13.98)  \\
 								& $t\!=\!20$		&  (1.62,2.69)  &   (0.63,8.31)   &	  (0.28,18.46)  \\ \hline
\multirow{4}{*}{\!$c\!=\!10$\!} & $t\!=\!0.5$ 		&  (3.01,0)		&	(3.13,1)      &	  (3.19,1)  \\
 								& $t\!=\!2$			&  (5.6,0)		&	(6.29,1.02)	  &	  (6.37,1.09) \\
 								& $t\!=\!5$			&  (6.47,0)		&	(6.77,1.25)	  &	  (5.79,2.08) \\
 								& $t\!=\!10$		&  (6.52,0)		&   (6.67,1.4)	  &	  (4.99,3.16) \\
 								& $t\!=\!20$		&  (6.52,0)	    &   (6.67,1.4)	  &	  (4.66,3.68)\\ \hline
\multirow{4}{*}{\!$c\!=\!15$\!} & $t\!=\!0.5$ 		&   -			&	(3.13,1)	  &	  (3.19,1)  \\
 								& $t\!=\!2$			&	-			&	(6.59,1)	  &	  (7.18,1)  \\
 								& $t\!=\!5$			&	-			&	(8.39,1)	  &	  (9.57,1.03)\\
 								& $t\!=\!10$		&	-			&   (8.68,1)	  &	  (10,1.07)  \\
 								& $t\!=\!20$		&   -		    &   (8.68,1)	  &	  (10.04,1.08) \\ \hline
\multirow{4}{*}{\!$c\!=\!20$\!} & $t\!=\!0.5$ 		& 	-			&	(3.13,0)	  &	  (3.19,1)  \\
 								& $t\!=\!2$			&	-			&	(6.59,0)      &	  (7.19,1)  \\
 								& $t\!=\!5$			&	-			&	(8.39,0)	  &	  (9.83,1)  \\
 								& $t\!=\!10$		&	-			&   (8.68,0)	  &	  (10.47,1)  \\
 								& $t\!=\!20$		&   -		    &   (8.69,0)	  &	  (10.52,1)  \\ \hline
\end{tabular}
\end{center}
\label{table1}
\end{table}
\begin{table}[t]
\caption{First moments under heterogeneous contacts.}
\vspace{-0.5em}
\begin{center}\small
\begin{tabular}{|l|l|c|c|c|}
\hline
\multicolumn{2}{|c|}{\!\big($E[I(t)]$, $E[R(t)]$\big)\!}  	        &  $N\!=\!10$   &	$N\!=\!20$    &   $N\!=\!40$      \\ \hline\hline
\multirow{4}{*}{\!$c\!=\!5$\!}	& $t\!=\!0.5$		&  (2.27,1)  	&   (2.51,1)	   &  (2.39,1)  \\
 								& $t\!=\!2$			&  (2.89,1.25)	&	(2.26,1.94)    &  (2.38,1.81)  \\
 								& $t\!=\!5$			&  (2.2,1.98)	&	(1.14,5.1)	   &  (1.19,5.15)  \\
 								& $t\!=\!10$		&  (1.89,2.3)	&	(0.67,8.19)	   &  (0.62,10.2) \\
 								& $t\!=\!20$		&  (1.83,2.37)	&	(0.48,9.71)	   &  (0.33,17.74) \\ \hline
\multirow{4}{*}{\!$c\!=\!10$\!} & $t\!=\!0.5$ 		&  (2.09,0)		&	(2.66,1)	   &  (2.51,1)   \\
 								& $t\!=\!2$			&  (4.03,0)		&	(5.59,1.01)    &  (5.34,1.01)   \\
 								& $t\!=\!5$			&  (5.23,0)		&	(6.52,1.25)    &  (6.36,1.29)  \\
 								& $t\!=\!10$		&  (5.48,0)		& 	(6.39,1.52)	   &  (6.12,1.77)  \\
 								& $t\!=\!20$		&  (5.49,0)		&	(6.33,1.6)	   &  (5.94,2.07) \\  \hline
\multirow{4}{*}{\!$c\!=\!15$\!} & $t\!=\!0.5$	 	& 	-			&	(2.66,1)	   &  (2.51,1)   \\
 								& $t\!=\!2$			&	-			&	(5.75,1)	   &  (5.48,1)   \\
 								& $t\!=\!5$			&	-			&	(7.88,1)       &  (7.79,1)  \\
 								& $t\!=\!10$		&	-			& 	(8.42,1)        &  (8.51,1.02)  \\
 								& $t\!=\!20$		&   -			&	(8.47,1)	   &  (8.6,1.02) \\ \hline
\multirow{4}{*}{\!$c\!=\!20$\!} & $t\!=\!0.5$ 		& 	-			&	(2.66,0)	   &  (2.51,1)	\\
 								& $t\!=\!2$			&	-			&	(5.75,0)	   &  (5.48,1)   \\
 								& $t\!=\!5$			&	-			&	(7.9,0)	   &  (7.84,1)  \\
 								& $t\!=\!10$		&	-			& 	(8.47,0)	  	   &  (8.61,1)  \\
 								& $t\!=\!20$		&   -			&	(8.52,0)	   &  (8.72,1) \\ \hline
\end{tabular}
\end{center}
\label{table2}
\vspace{-0.5em}
\end{table}

\subsection{Impact of Retrial Rates at Different Time Points}
\begin{figure}[!t]
    \centering
    \begin{subfigure}{0.48\textwidth}
        \includegraphics[width=\textwidth]{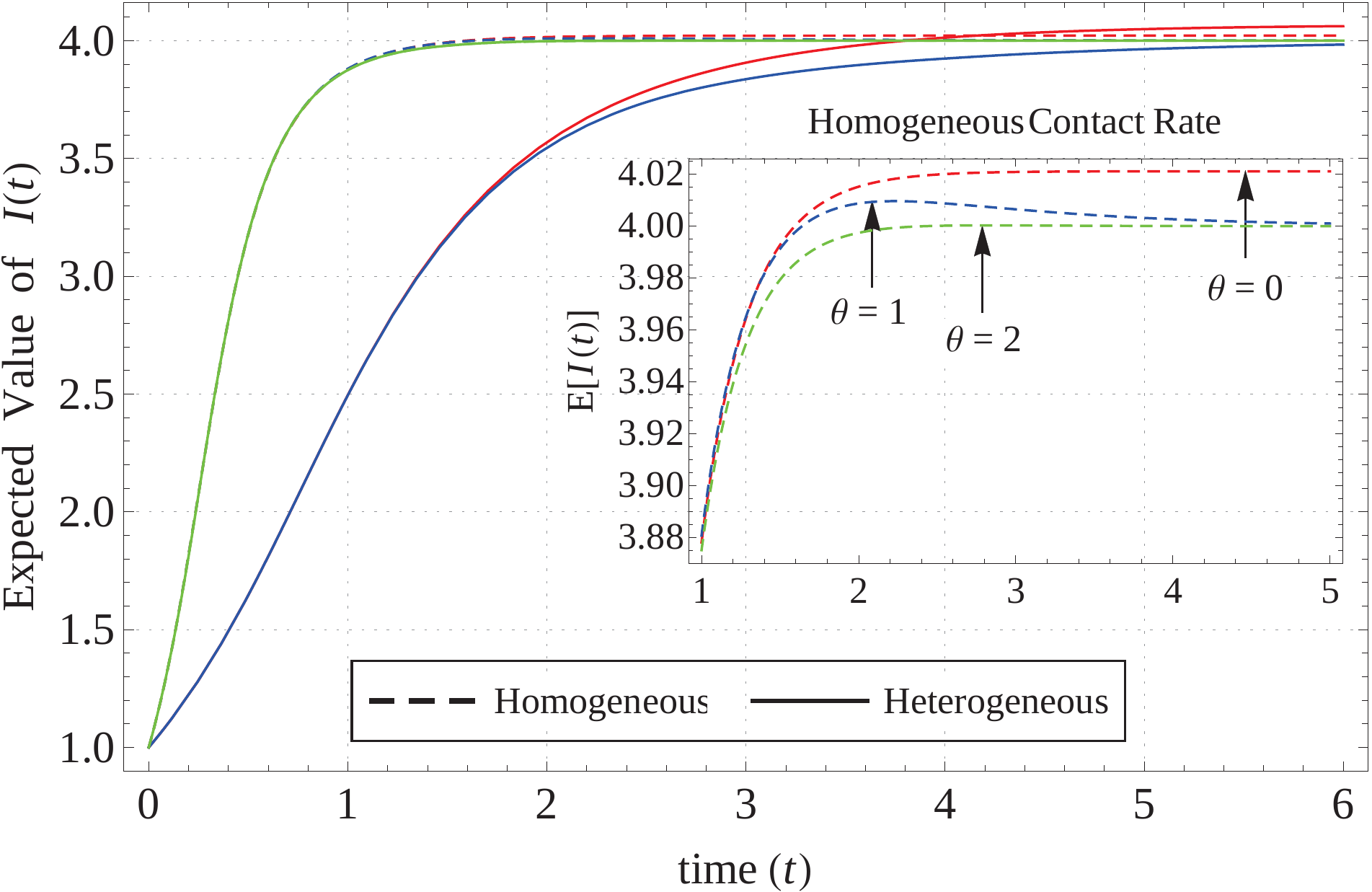}
        \caption{Expected number of infectives under recovery.}
        \label{fig5a}
        \vspace{0.5em}
    \end{subfigure}
      
    \begin{subfigure}{0.48\textwidth}
        \includegraphics[width=\textwidth]{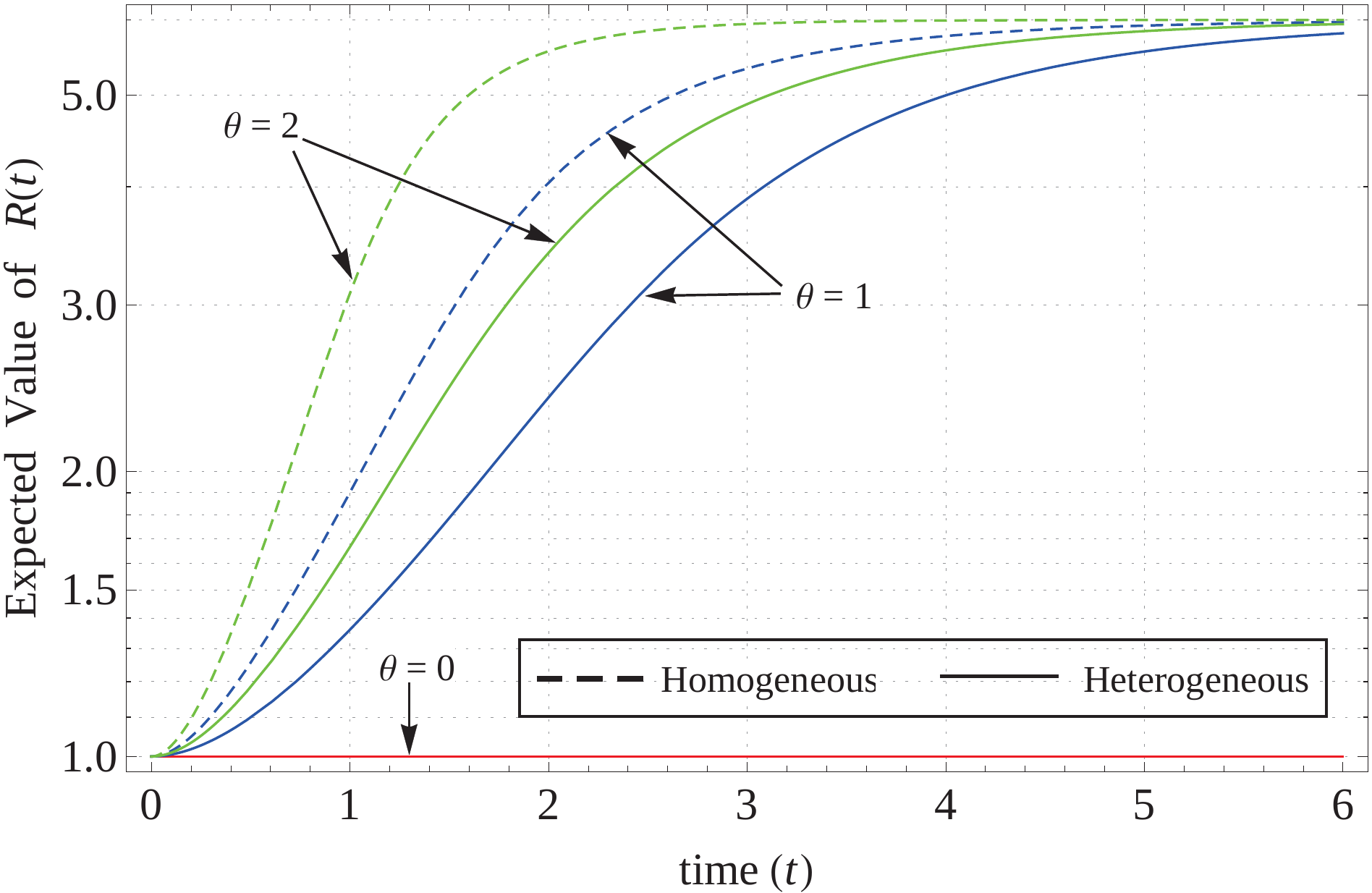}
        \caption{Expected number of infectives in orbit.}
        \label{fig5b}
    \end{subfigure}
    \caption{Expected values under varying retrial rates for $N\!=\!20$, $c\!=\!8$, $\alpha\!=\!5$, $\mu\!=\!1$, $\beta_{k,l}\!=\!1$, and $p_{0,0}(0)\!=\!1$.}
    \vspace{-0.5em}
    \label{fig5}
\end{figure}

The evolution of expected values of $I(t)$ and $R(t)$ in terms of the retrial rate $\theta$ are shown in Fig.~\ref{fig5} for $(N,c)=(20,8)$ and $\mu\!=\!1$. Fig.~\ref{fig5a} depicts that the difference in the average number of recovering nodes is very small for given $\theta$ values in both contact types. However, the transient state for the homogeneous case lasts for a much shorter time than its heterogeneous counterpart. Additionally, for $\theta\!=\!0$, $E[I(t)]$ reaches a slightly higher steady-state value than that for $\theta\!>\!0$. Intuitively, for $\theta\!=\!0$, the rate at which nodes get infected in the reduced SI model with limited recovery units solely depends on the number of busy recovery units. The impact of retrial on the number of orbital nodes is given in Fig.~\ref{fig5b}. While the model with heterogeneities follows an identical trend, it reveals lower values for $E[I(t)]$ and $E[R(t)]$. The rationale behind such variation is the high pace of infectivity caused by the subsumed heterogeneous infection rates.

\section{Conclusions and Future Works}
\label{sec5}
We presented an exact Markov chain model for the SI epidemic process incorporated with the retrial attempts of infected nodes instigated by a recovery policy. Motivated by the significance of transient behavior statistics in practical online scanning services, we numerically obtain the marginal probability distributions of the number of recovering and orbital infectives and their corresponding moments for the proposed model under homogeneous contacts. Accounting for the possibility of infections caused by external and internal sources, the study was further extended to unravel the impact of infection heterogeneity on the network characteristics.

We believe utilizing the retrial notion under resource constraints is very promising and novel in epidemic modeling thus, breeding several open problems. A prospective follow-up on this work is a reasonably accurate mean-field approximation geared for asymptotic steady-state distribution analysis. Moreover, numerical results for cases of interest, namely large $N$ and $c$ values as well as quasi-stationary distribution of the number of infected nodes can be further investigated. From the viewpoint of  resource budgeting and optimization, statistics obtained in the transient regime can be used to find the optimal number of recovery units required to prevent an endemic in its early stages.

\section*{Acknowledgment}
This research was a part of the project titled ``Development of an Automated Fish-counter System and Measurement of Underwater Farming-fish'', funded by the Ministry of Oceans and Fisheries, South Korea.

\end{document}